\documentclass{article}
\usepackage{spconf,amsmath,graphicx,booktabs,url}
\usepackage{setspace}
\usepackage{multirow}
\usepackage[colorlinks=true, urlcolor=blue]{hyperref}
\usepackage{xcolor}
\hypersetup{
    colorlinks,
    linkcolor={black!50!black},
    citecolor={black!50!black},
    urlcolor={blue!80!black}
}

\title{Advancing Speech Translation: A Corpus of Mandarin-English Conversational Telephone Speech}
\name{Shannon Wotherspoon, William Hartmann, Matthew Snover}
\email{\{shannon.wotherspoon, william.hartmann, matt.snover\}@rtx.com}
\address{Raytheon BBN, Cambridge, MA-02138, USA}

\begin{document}
\maketitle

\begin{abstract}
     This paper introduces a set of English translations for a 123-hour subset of the CallHome Mandarin Chinese data and the HKUST Mandarin Telephone Speech data for the task of speech translation. Paired source-language speech and target-language text is essential for training end-to-end speech translation systems and can provide substantial performance improvements for cascaded systems as well, relative to training on more widely available text data sets. We demonstrate that fine-tuning a general-purpose translation model to our Mandarin-English conversational telephone speech training set improves target-domain BLEU by more than 8 points, highlighting the importance of matched training data.
\end{abstract}


\section{Introduction}
\label{sec:intro}
The availability of quality in-domain training data plays an indispensable role in the development of machine translation (MT) systems. While many data sets exist for text-based translation tasks, there is a scarcity of matched training data for the conversational speech domain. This paper introduces a new corpus of conversational speech translation data for Mandarin-English translation, addressing a critical gap in resources and underscoring the importance of domain-specific data in advancing the state-of-the-art in speech translation. 

\section{Data}
\label{sec:data}
In this section, we provide an overview of the data, including details of its collection and translation. The corpus consists of 123.5 hours of Mandarin conversational telephone speech (CTS) data, sourced from the CallHome Mandarin Chinese Speech\footnote{\href{https://catalog.ldc.upenn.edu/LDC96S34}{https://catalog.ldc.upenn.edu/LDC96S34} (CallHome speech)} \footnote{\href{https://catalog.ldc.upenn.edu/LDC96T16}{https://catalog.ldc.upenn.edu/LDC96T16} (CallHome transcripts)}\cite{CallHome_Speech, CallHome_Transcripts} and HKUST Mandarin Telephone Speech\footnote{\href{https://catalog.ldc.upenn.edu/LDC2005S15}{https://catalog.ldc.upenn.edu/LDC2005S15} (HKUST speech)}\footnote{\href{https://catalog.ldc.upenn.edu/LDC2005T32}{https://catalog.ldc.upenn.edu/LDC2005T32} (HKUST transcripts)} \cite{HKUST_Speech, HKUST_Transcripts} data sets. The CallHome portion of the data set contains 33.5 hours of speech from 242 unscripted telephone conversations between native Mandarin speakers. Most conversations are between close friends or relatives. The HKUST portion of the data set contains 90 hours of speech from 1,124 conversations between Mandarin speakers in Mainland China. Speakers in the HKUST subset are not necessarily native speakers of standard Mandarin, and may have regional accents. Furthermore, unlike the CallHome data, HKUST speakers did not necessarily know each other beforehand, and were given Fisher-style \cite{Fisher} topic prompts to encourage conversation.

The data are split into train, development, and test sets. The two development sets (dev1 and dev2) as well as the test set consist entirely of CallHome conversations, whereas the train set is a mix of CallHome and HKUST. The data breakdown by subset is detailed in Table \ref{tab:data}.

\begin{table}[th]
  \caption{Train, Development, and Test Sets}
  \label{tab:data}
  \centering
  \begin{tabular}{ l l r r r r }
    \toprule
    & & \textbf{train} & \textbf{dev1} & \textbf{dev2} & \textbf{test} \\
    Hours & HKUST & 90.1 & 0 & 0 & 0 \\
    & CallHome & 20.5 & 4.0 & 4.5 & 4.4 \\ 
    & \textbf{Total} & \textbf{110.6} & \textbf{4.0} & \textbf{4.5} & \textbf{4.4} \\
    \midrule
    Conversations & HKUST & 1122 & 0 & 0 & 0 \\
    & CallHome & 114 & 8 & 11 & 9 \\  
    & \textbf{Total} & \textbf{1236} & \textbf{8} & \textbf{11} & \textbf{9} \\
    \midrule
    Utterances & HKUST & 103042 & 0 & 0 & 0 \\
    & CallHome & 36842 & 6964 & 7819 & 7193 \\  
    & \textbf{Total} & \textbf{139884} & \textbf{6964} & \textbf{7819} & \textbf{7193} \\
    \bottomrule
   \end{tabular}
\end{table}

\subsection{Speech Data Translation}

The primary contribution of this paper is a set of English translations for the Mandarin speech data, enabling this corpus to be used to build speech translation systems\footnote{Contact the authors for access to the translations.}. The translations were provided by Mandarin-English bilingual annotators through Appen. The annotators were given the Mandarin transcripts and asked to translate on an utterance-by-utterance basis. Annotators were not able to listen to the corresponding audio, but did have access to surrounding utterances for context, though utterances with identical text were translated only once, regardless of how often they occurred in the corpus. Translators were instructed to preserve any disfluencies, hesitations, or code-switching present in the data. For quality assurance, the translations provided by Appen underwent multiple iterations of feedback with coordinators at Raytheon BBN, and translations were sourced from multiple annotators.

\section{Experiments}
In this section, we present results from cascade speech translation systems, where the decoder output of an Automatic Speech Recognition (ASR) system is used as input to an MT system. For our ASR model, we use a hybrid TDNN-LSTM \cite{TDNN_LSTM} trained with Raytheon BBN's speech recognition system, Sage \cite{Sage}, which makes use of the Kaldi speech recognition toolkit \cite{Kaldi}. The model is trained on the Mandarin CTS train set, plus an additional 137 hours of Mandarin ASR-only data from the HKUST corpus. The Word Error Rate (WER) of the ASR model on the Mandarin CTS test set is 26.7. For our MT models, we use the No Language Left Behind (NLLB) model \cite{NLLB} both with and without fine-tuning to the Mandarin-English CTS train set. Results are presented on the Mandarin-English CTS test set in Table \ref{tab:results}.

The results in Table \ref{tab:results} highlight the importance of matched training data when building conversational speech translation systems. Without fine-tuning, the NLLB model, a general-purpose translation model trained largely on text data, performs poorly on the CTS test set, with a BLEU score of 5.98. However, after fine-tuning to the CTS train set, BLEU improves by 137\% relative. 

\begin{table}[th]
  \caption{Speech Translation Results}
  \label{tab:results}
  \centering
  \begin{tabular}{ l l r }
    \toprule
    \textbf{ASR System} & \textbf{MT System} & \textbf{BLEU} \\
    \midrule
    Hybrid TDNN-LSTM & NLLB & 5.98  \\
    Hybrid TDNN-LSTM & NLLB ft CTS Train & 14.16 \\
    \bottomrule
  \end{tabular}
\end{table}

\section{Conclusion}

This paper has shown the importance of domain-specific, matched training data for building conversational speech translation systems. Our empirical results highlight the performance gap between general-purpose translation models such as NLLB and a model tuned specifically to our target domain. While general-purpose models may suffice for some domains, they fall short for others, such as the Mandarin conversational speech domain, where BLEU scores from general-purpose MT models are so low as to be unusable. In order to build quality conversational speech translation models, we have demonstrated that matched training data is essential. The corpus introduced in this paper provides a critical resource for the research and development of such systems.


\vfill\pagebreak

\bibliographystyle{IEEEbib}
{\footnotesize
\bibliography{refs}}

\begin{thebibliography}{1}

\bibitem{CallHome_Speech}
Alexandra Canavan and George Zipperlen,
\newblock ``{CALLHOME Mandarin Chinese Speech LDC96S34},'' Web Download. Philadelphia: Linguistic Data Consortium, 1996.

\bibitem{CallHome_Transcripts}
Barbara Wheatley,
\newblock ``{CALLHOME Mandarin Chinese Transcripts LDC96T16},'' Web Download. Philadelphia: Linguistic Data Consortium, 1996.

\bibitem{HKUST_Speech}
Pascale Fung, Shudong Huang, and David Graff,
\newblock ``{HKUST Mandarin Telephone Speech, Part 1 LDC2005S15},'' Web Download. Philadelphia: Linguistic Data Consortium, 2005.

\bibitem{HKUST_Transcripts}
Pascale Fung, Shudong Huang, and David Graff,
\newblock ``{HKUST Mandarin Telephone Transcript Data, Part 1 LDC2005T32},'' Web Download. Philadelphia: Linguistic Data Consortium, 2005.

\bibitem{Fisher}
Christopher Cieri et~al.,
\newblock ``{Fisher English Training Speech Part 1 Transcripts LDC2004T19},'' Web Download. Philadelphia: Linguistic Data Consortium, 2004.

\bibitem{TDNN_LSTM}
Vijayaditya Peddinti, Yiming Wang, Daniel Povey, and Sanjeev Khudanpur,
\newblock ``Low latency acoustic modeling using temporal convolution and {LSTMs},''
\newblock {\em IEEE Signal Processing Letters}, vol. 25, no. 3, pp. 373--377, 2017.

\bibitem{Sage}
Roger Hsiao, Ralf Meermeier, Tim Ng, Zhongqiang Huang, Maxwell Jordan, Enoch Kan, Tanel Alum{\"a}e, Jan Silovsk{\`y}, William Hartmann, Francis Keith, et~al.,
\newblock ``Sage: The new {BBN} speech processing platform.,''
\newblock in {\em Proc. INTERSPEECH 2016}, San Francisco, California, USA, 2016, pp. 3022--3026.

\bibitem{Kaldi}
Daniel Povey, Arnab Ghoshal, Gilles Boulianne, Lukas Burget, Ondrej Glembek, Nagendra Goel, Mirko Hannemann, Petr Motlicek, Yanmin Qian, Petr Schwarz, et~al.,
\newblock ``The {Kaldi} speech recognition toolkit,''
\newblock in {\em {IEEE} 2011 workshop on automatic speech recognition and understanding}. IEEE Signal Processing Society, 2011.

\bibitem{NLLB}
Marta~R Costa-juss{\`a}, James Cross, Onur {\c{C}}elebi, Maha Elbayad, Kenneth Heafield, Kevin Heffernan, Elahe Kalbassi, Janice Lam, Daniel Licht, Jean Maillard, et~al.,
\newblock ``No language left behind: Scaling human-centered machine translation,''
\newblock {\em arXiv preprint arXiv:2207.04672}, 2022.

\end{thebibliography}

\end{document}